\def \cm{~\rm{cm}}
\def \s{~\rm{s}}
\def \km{~\rm{km}}

\def \K{~\rm{K}}
\def \g{~\rm{g}}

\def \erg{~\rm{erg}}

\def \yr{~\rm{yr}}

\def \kpc{~\rm{kpc}}

\documentclass[12pt,preprint]{aastex}
\begin{document}

\title{A POSSIBLE HIDDEN POPULATION OF SPHERICAL PLANETARY NEBULAE}

\author{Noam Soker and Eyal Subag}
\altaffiltext{1} {Department of Physics, Technion$-$Israel
Institute of Technology, Haifa 32000 Israel;
soker@physics.technion.ac.il.}

\begin{abstract}
We argue that relative to non-spherical planetary nebulae (PNs),
spherical PNs are about an order of magnitude less
likely to be detected, at distances of several kiloparsecs.
Noting the structure similarity of halos around non-spherical PNs
to that of observed spherical PNs, we assume that most unobserved
spherical PNs are also similar in structure to the spherical halos
around non-spherical PNs.
The fraction of non-spherical PNs with detected spherical halos
around them, taken from a recent study, leads us to the claim
of a large (relative to that of non-spherical PNs) hidden
population of spherical PNs in the visible band.
Building a toy model for the luminosity evolution of PNs, we show
that the claimed detection fraction of spherical PNs based on halos
around non-spherical PNs, is compatible with observational sensitivities.
We use this result to update earlier studies on the different
PN shaping routes in the binary model.
We estimate that $\sim 30 \%$ of all PNs are spherical, namely, their progenitors
did not interact with any binary companion. This fraction is to be compared
with the $\sim 3 \%$ fraction of observed spherical PNs among all observed PNs.
From all PNs, $\sim 15 \%$ owe their moderate elliptical shape to the
interaction of their progenitors with planets, while $\sim 55 \%$ of all PNs
owe their elliptical or bipolar shapes to the interaction of their
progenitors with stellar companions.
\end{abstract}

\keywords{planetary nebulae: general --- stars: AGB and post-AGB ---
stars: mass loss}

\section{INTRODUCTION}
Planetary nebulae (PNs) are the descendants of intermediate mass
stars (initial masses $\sim 1-8 M_\odot$).
Much of the star's initial mass is expelled during the $\sim 10^5 \yr$
that the star spends on the upper asymptotic giant branch (AGB).
This AGB star wind usually consists of a more or less spherically
symmetric outflow at rates of
$\sim 10^{-7}-10^{-5} M_\odot \yr ^{-1}$.
Most PNs, though, possess a global axisymmetrical rather than a spherical
structure in their inner region.
This is most evidence in the spherical halos observed around many
axisymmetrical PNs (e.g., Corradi et al. 2003, 2004).
In axisymmetrical we include also point-symmetric PNs, which are defined
as PNs having more than one axisymmetric substructure, but the different
symmetry axes intersect in a common point. This class includes also the
quadrupolar PNe defined by Manchado et al. (1996b).
These axisymmetrical PNs with spherical faint halos suggest that only
close to the termination of the AGB, and/or later in the early post-AGB
phase, the outflow acquires the axisymmetrical structure.
The axisymmetrical inner structure of these PNs are much brighter than
the spherical halo, implying that the mass loss rate which formed these
parts was much higher than the regular AGB wind.
As pointed out in earlier papers, e.g., Soker (2000; 2002), there
is a positive correlation between the onset of this
final intensive wind (FIW; termed also superwind)
and the transition from spherical to axisymmetrical mass loss.
This correlation is well explained by the binary model for the formation
of non-spherical PNs (Soker 2000; 2004), where the binary can be  a
stellar or a substellar object.

Different classification schemes have been suggested over the years of
the different morphological PN types, e.g., Manchado et al. (1996a ; 2000).
We will follow Soker (1997) in a classification scheme which is
motivated in part by theoretical considerations, and which
contains four basic types of structure.
\begin{enumerate}
\item {\it Spherical PNs.} These are defined as PNs which shows no
axisymmetrical structure at all.
The properties of these PNs were studied by Soker (2002).
One prominent property is that these PNs look like the
spherical halos detected around some elliptical PNs, in
that there is no indication for a FIW (superwind).
Note that the definition of spherical PNs assumed here
is quite different from that of Manchado et al.\ (2000).
Manchado et al.\ find that according to their definition
$\sim 25 \%$ of all observed PNs are spherical, while Soker
(1997) finds that only $\sim 3-4 \%$ of all observed PNs are spherical.
This difference in definition is summarized on one side
by Soker (2002), and on the other side by Stanghellini
et al.\ (2002).
There are 14 such PNs (those with no signature of a FIW) in the list
of 18 PNs in Table 2 of Soker (1997).
\item {\it Elliptical PNs.}
These are PNe having a large scale elliptical shape, with a shallow density
variation from equator to poles. They may possess jets.
Examples are most PNs listed in Table 5 of Soker (1997).
\item {\it Extreme-elliptical PNs.}
These are defined as PNe having a large equatorial concentration of mass, e.g.,
a ring, but no, or only small, lobes. They may possess jets.
Examples are most PNs listed in Table 4 of Soker (1997).
\item {\it Bipolar PNs.}
Bipolar (also called ``bilobal'' and ``butterfly'') PNs are defined
(Schwarz et al. 1992) as axially symmetric PNs
having two lobes with an equatorial waist between them.
Examples are in Corradi \& Schwarz (1995) and most PNs listed in
Table 3 of Soker (1997).
\end{enumerate}

Efforts have been made over the last decade to estimate the
fraction of PNs belonging to each group, and from that to better
understand the shaping of PNs.
The major question is whether the formation of axisymmetrical PNs
requires the presence of a binary companion, or whether a single
star can do the job.
We here follow earlier works of Soker (1997, 2001a,b) connecting
different axisymmetrical structures to different types of
interaction with stellar and substellar companions.
This is summarized in section 2, where an updated estimate of the
fraction of PNs shaped by planets is given.
Motivated by new statistical works on extra-solar planets, and
the new work of Corradi et al. (2003) on halos around PNs, we
set a goal to update the statistics of the different classes
of PNs and their relation to the progenitor type.
Using the results of Corradi et al.\ (2003) with our
two basic assumptions, we estimate in section 3 the detection fraction
of spherical PNs relative to that of non-spherical PNs.
In section 4 we present a phenomenological toy model, that shows
that the present detection limits of faint PNs might be compatible
with the presence of a large hidden spherical PN population.
 By hidden population we refer to population of PNs which are
below the threshold for being discovered by most visible-light surveys.
We limit ourselves to the visible band because all large morphological
samples to which shaping theories are compared are based on visible-light images.
Some PNs in the visible-hidden population might be still detected
in the IR, but this will not affect our claims regarding the present
morphological classification to which theories are compared.
In section 5 we discuss and summarize our main results and predictions.

\section{STATISTICS AND BINARY INTERACTION}
Only $\sim 60 \%$ of all systems are binary stellar systems
(Duquennoy \& Mayor 1991), and only in a fraction of these the
companion is close enough to influence the mass loss from the
PN progenitor.
The fraction of observed axisymmetrical PNs, on the other hand,
is much larger. Soker (1997) finds from classifying the morphology
of 458 PNs this fraction to be $96 \%$,
namely, $~4 \%$ of all observed PNs are spherical.
Soker (1997) considered selection effects, and took the true
fraction of spherical PNs to be $10 \%$.
The shaping of the non-spherical (axisymmetrical) PNs
was attributed to three basic types of interaction as follows:
(1)  Progenitors that interact with close stellar companions that avoided
a common envelope phase, or enter to a common envelope at a very late
stage of the FIW phase, and form mainly bipolar PNs;
$\sim 11\%$ of all PNs.
(2) Progenitors that interact with stellar companions via a common
 envelope phase and form mainly extreme elliptical PNs;
 $\sim 23 \%$ of all PNs.
(3) Progenitors that interact with substellar, i.e., planet and brown
dwarf, companions via a common envelope phase and form elliptical
PNs (with moderate or small departure from sphericity);
$\sim 56 \%$ of all PNs.

The interaction with the substellar companion scenario had
three major predictions.
\newline
($i$) Many planets will be more massive and closer to their parent
stars than Jupiter is (Soker 1994, 1996).
\newline
($ii$) For many stars to engulf a planet at their late evolutionary
stages with a high probability, several substellar objects must be
present in most of the systems (Soker 1996).
\newline
($iii$) About half of all progenitors of PNe have
planetary systems with massive planets (Soker 1997).
This prediction was based on the notion that
singly evolved stars rotate too slow when they become AGB
stars, and hence have spherical mass loss.
\newline
The first two predictions were made before extrasolar planets
were found, and we adopt them here as well.

In Soker (1996) massive planets were regarded as planets of mass
$M_p \ga M_J$, where $M_J$ is Jupiter mass, while in Soker (2001b)
the minimum allowed mass to cause departure from spherical mass
loss was reduced to $M_p \ga 0.01 M_J$.
This reduced minimum planetary mass was motivated by the finding that
only $\sim 5 \%$ of solar-like stars have Jupiter like planets around
them (Marcy \& Butler 2000), and is based on non-linear effects in the
AGB envelope, such that slow rotation caused by low mass planets can
lead to axisymmetrical mass loss.
These effects might be excitation of non-radial pulsations,
or local (but not global) dynamo formation of magnetic cool spots;
the lower temperature above the cool spots enhanced dust formation,
hence mass loss rate.
Other non-linear processes based on locally strong
magnetic fields are possible.
For example, leakage of p-modes (sound waves) were recently proposed
as the mechanism for the formation of chromospheric spicules in the Sun
(De Pontieu et al.\ 2004);
spicules are dynamic jets propelled upwards from the solar surface at
much below escape speed.
In AGB stars such process might enhance dust formation, and hence
increase mass loss rate. If the magnetic spots are concentrated in the
equatorial plane (Soker 2001b), the enhanced mass loss rate will
come with axisymmetric mass loss geometry as well.

Soker (2001a) studied the formation of (moderate) elliptical PNs by
stellar companions orbiting the progenitor at large orbital separations.
The companion accretes mass, forms an accretion disk, and blows two jets.
This process was found to shape $\sim 5-20 \%$ of axisymmetrical PNs,
hence reducing the percentage of PN shaped by planets
from $\sim 50 \%$ to $\sim 30 - 45 \%$.
The estimated numbers, in the stellar and substellar shaping model,
in 2001 (Soker 2001a,b) were as follows (Table 1; for the different types of
    the binary interaction see Soker 2004):
(1) AGB stars which interact with a close stellar companion that avoided
    a common envelope: $\sim 15 \%$.
(2) AGB stars which interacted with a stellar companion via a common
    envelope: $\sim 25 \%$ (Yungelson et al. 1993; Han et al.\ 1995).
(3) AGB stars which interacted with a substellar (mainly planets)
    companion massive enough to cause observable axisymmetrical
    structure: $\sim 35 \%$. (This is found to be
       $\sim 15-20 \%$ in the present paper).
(4) AGB stars which interacted with a wide stellar companion:
      $\sim 15 \%$.
(5) AGB stars which did not interact with any companion and formed
   spherical PNs: $\sim 10 \%$. (This is found to be
       $\sim 25-30 \%$ in the present paper).

As it turned out in recent years, the previously estimated
percentage of PN progenitors interacting with planets is
higher than the observed percentage of stars hosting planets.
Santos et al. (2004), for example, find that $\sim 3 \%$ of
stars with solar-metallicity have planets.
The number rises to $\sim 25-30 \%$ for stars with twice
solar metallicity ([Fe/H]$>0.3$).
Lineweaver \& Grether (2003) extrapolate from the detected parameters
space of planetary system properties to that below detection
sensitivity, and conclude that the real number of
planetary system is much larger.
The fraction of solar-like stars having planets of
$M_p>0.1 M_J$ and that are close enough to interact
with their parent stars, is $\ga 20 \%$.
Over all, therefore, $\sim 25 \%$ of stars have planets
that are massive and close enough to influence their mass loss
process; a fraction higher than the one deduced by Livio \& Soker
(2002) who used a much stronger constraint on planets to
influence their parent stars mass loss process.

Two effects should be considered here.
First, if most PNs are formed from more massive than the Sun stars,
which on average are more metal rich, then the percentage of
progenitors interacting with planet increases.
Second, as note by Soker (2001b), most known Sun-like stars
that have planets around them will not form PNs at all.
Their planet(s) will deposit enough orbital angular momentum and
energy to cause most, or even all, of the envelope of these
stars to be lost already on the red giant branch.
These stars will not evolve along the AGB, or evolved only along
the early AGB. In any case, they will have too low mass
envelope to form a detectable PN.
This reduces the fraction of PN progenitors interacting with
planets.

 The main conclusion form this discussion is that with present
statistics of extrasolar planets, the percentage of PN progenitors
that are expected to be influenced by planets is much less
than $\sim 35 \%$, and it is more like $\sim 15-20 \%$.
The percentages given above for the different interaction types
with stellar companions are based on a more robust stellar population
studies, and hence are more secure.
In addition, recent observations suggest that the central star
of many PNs have companions (De Marco et al. 2004; Hillwig 2004;
Pollaco 2004).
The lower observe percentage of stars hosting planetary systems
than the percentage given in previous studies motivate us to examine
the possibility that the true fraction of spherical PNs is much
larger than the one observed.

\section{SPHERICAL HALOS AND MASS LOSS HISTORY}

The FIW (superwind) at the end of the AGB is preceded by lower mass
loss rate episodes, which in most cases known are spherical.
Therefore, from theoretical considerations we expect all PNs to
have a faint halo around them. However, the detection fraction
of faint halos is much lower.
We also expect that spherical PNs will be PNs formed by
AGB stars which did not experienced the FIW (superwind).
We therefore take the following to be the basic, quite
reasonable, assumption (Soker 2002):
{\it Spherical PNs are similar in structure to the faint large
halo of elliptical and extreme elliptical PNs.}
For example, the spherical PN A30 is similar to the halo of NGC 6826.

Corradi et al. (2003) list the 49 PNs known to have halos. We
estimate the total halo detection rate as follows. For the sample
of all well resolved PNs we use the statistics of Soker (1997),
were 4 classes are defined (previous sections). We do not consider
the spherical PNs (Table 2 in Soker 1997), and the bipolar PNs
having large lobes and a narrow waist, as in general they were not
observed by Corradi et al. (2003). The rest are 275 elliptical PNs
and 113 extreme elliptical PNs.
Around these PNs detection of halos is easier
  than around the bright and highly non-spherical
bipolar PNs.
Therefore, the detection fraction
of halos is $\chi \sim 49/(275+113)=0.13$; if we include all PNs
studied by Soker (1997), the halo detection fraction is $\chi \sim
49/458=0.11$. The random detection fraction is even lower, because
Corradi et al. (2003) were aiming at specific PNs in their search
for halos.

If we use our basic assumption in this section, that spherical
PNs are similar to faint halos around elliptical and
extreme elliptical PNs, we deduce that the detection fraction
of spherical PNs is $\chi \sim 0.12$ times that of non-spherical PNs.
This implies that the fraction of spherical PNs is
$\sim 20-30 \%$, rather than the observed $\sim 3 \%$.

  We stress that the detection fraction of PNs with halos
as taken from Corradi et al.\ (2003) is an upper limit.
This is because the search was aiming at PNs likely to have
halos, whereas the 388 PNs were collected by random detection.
Therefore, the real fraction of spherical PNs can be
higher than $\sim 20-30 \%$.
The number of PNe known to have halos prior to the work
of Corradi et al.\ (2003) is $~25$
(R. Corradi, private communication 2005).
These amount to $\sim 6 \%$ of the 388 PNs, rather than the $\sim 12 \%$
upper limit taken from Corradi et al.\ (2003).
If all PNs are expected to have halos, then the detection fraction
of non-spherical PNs is $\sim 16$ times that of spherical PNs.
Then the fraction of spherical PNs is $\sim 35 \%$.
If, on the other hand, only $\sim 50-60 \%$ of PNs have halo
(R. Corradi, private communication 2005), then the
fraction of spherical PNs is again $\sim 25 \%$.

Halos are detected by their surface brightness.
In the next section we consider the detection fraction
from the luminosity of PNs.

\section{ESTIMATING THE UNDETECTED POPULATION BY LUMINOSITY SEARCH}
The goal of the present section is to show that it is quite
possible that a hidden population of spherical PNs exists
in detection and surveys based on luminosity rather than surface
brightness.
This will give a lower limit on the hidden spherical
population, as spherical PNs are even more difficult to detect
relative to non-spherical PNs via surface brightness.
We will build a very simple toy model, then  choose typical values for
the different physical variables, and show that many spherical PNs may
have escape detection.
  This is simply because the spherical
halo are very dim, hence relative to elliptical and bipolar PNs,
spherical PNs are much less likely to be detected.
We stress that we do not prove the existence of such a population.
We only show, in line with the discussion in the previous section,
that because the halos expected to be present around many
non-spherical PNs are not detected, many (if not most) spherical
PNs may escape detection.

 Our goal is to update the statistics of the different
morphological classes of PNs and their relation to the progenitor
type in the binary progenitor theory.
Since all large samples of PNs which are classified by morphology and
to which theory is compared are visible-light samples, this study is limited
to the discovery of PNs in the visible band.
We note however that many PNs are bright in the IR, as was already
found by IRAS, with typical flux values of $\sim 1$~Jy at $12 \mu$m,
and $\sim 10-100$~Jy at $60 \mu$m (Pottasch et al.\ 1984).
The IR emission can be accounted for with models of
PN evolution (Volk 1992).
PNs are also detected in the near IR, e.g., by the 2MASS all-sky
infrared survey (Ramos Larios \& Phillips 2005).
In most cases the near-IR emission is dominated by
the central star continuum, but extended PN are also found
(Ramos Larios \& Phillips 2005).
Although relatively bright in the IR, most PNs have been discovered by
visible-light emission.
Some PNs were discovered by their IR emission (e.g., Beer \& Vaughan 1999),
but even then [OIII] emission is required for definite identification.
When large enough catalogues of well resolved PNs in the IR exists,
the same type of studied performed here for visible-light classification of PNs
can be done for IR classification.
We here note that a hidden population of spherical PNs is expected
also in samples of PNs discovered by their IR emission, because
bipolar PNs, or those having symbiotic outflows, are distinguished
from other PNs (e.g., M2-9; Ramos Larios \& Phillips 2005;
Schmeja \& Kimeswenger 2001).
Some bipolar young PNs and proto-PNs are particularly bright in the IR,
e.g., the Red Rectangle (Miyata 2004), also suggesting that a hidden
population will exist in future morphological catalogues based on
PNs discovered by their IR emission.
{{{{{ Bright PNs in the near-IR, as another example, were formed by progenitors
with high mass loss rates (Hora et al. 1999).
Such progenitors tend to form bipolar PNs.
Indeed, most bright PNs in the near-IR are bipolar PNs
(e.g., Latter et al. 1995, 2000; Hora \& Latter 1996; Hora et al. 1999, 2004;
Kastner et al. 2001, 2002; Meakin et al. 2003).
The positive correlation between near-IR emission and bipolar morphology
(e.g., Latter et al. 1995) shows that in the near-IR an even larger relative
hidden population of spherical PNs might exist,
adding support to the claim in the present paper. }}}}}

\subsection{The Toy Model}
Considering the other uncertainties in estimating the progenitor
population of PNs, we build a toy model to estimate the fraction
of the undetected spherical PN population.
We make the following assumptions.
\begin{enumerate}
\item Spherical PNs are similar in structure to the faint large
halo of elliptical and extreme elliptical PNs.
  This assumption is based on the results of Soker (2002), who argues
that the main difference between spherical and non-spherical PNs with halos,
is that the progenitors of spherical PNs did not go through the FIW
(superwind) phase.
Namely, the evolution of spherical, elliptical, and extreme-elliptical
is similar until the onset of a FIW in the final stages of the
AGB phase of the progenitors of elliptical and extreme-elliptical PNs.
This crucial assumption for the model seems to be strongly supported
by observations,
mainly by the similar structure of spherical PNs to that of the structure
of the halo of elliptical and extreme-elliptical PNs (Soker 2000; 2002).
\item We assume that for a PN to be discovered, its visible energy flux
(observed luminosity) should be above some threshold $F_0$ at a considered
line, say $H\beta$.
  Had we used limit on the surface brightness rather than on the
luminosity, the relative number of undetected spherical PNs compared
with undetected non-spherical-PNs would be larger, strengthening
our results (see last paragraph of section 4.3).
\item We assume that PNs are composed of three parts formed from
three mass loss phases.
A regular AGB wind with a constant mass loss rate of $\dot M_h$ formed
the halo.   It started at time $t_h$ before present, and ceased at
time $t_s$ before present.
A later FIW with mass loss rate of $\dot M_s$ forms the dense shell.
It started at time $t_s$ before present, and ceased at time $t_r$ before present.
Both winds expand with the same speed $v$.
Finally, a faster wind, but of low density, accelerates the inner part of the
shell. This forms a dense and thin shell, termed rim
(Frank et al. 1990).
\item We assume $t_h \gg t_s$, i.e., the FIW (superwind) phase is relatively
short.
\item We neglect the time interval between end of the AGB, when we set $t_r=0$,
and the time the PN turns on, namely, when the central star ionizes most
of the nebula. This is justified as PNs live for tens of thousands of year,
while the post-AGB phase lasts few thousands years or less.
We also assume that the life time of all PNs is $t_L$, with no
dependence on the morphological type (no dependence on $\eta$;
see eq. \ref{eta1}).
\item We neglect the contribution of the rim to the visible
total luminosity, as well as of the inner tenuous bubble.
We also neglect the change in density structure resulting from the
propagation of the ionization and shocks fronts (Perinotto et al. 2004).
\item We assume a spherical PN structure in calculating the luminosity.
\item All PNs, with no dependence on morphological type, have the same
$\dot M_h$.
Following the non-detection of halos around many PNs (see previous section),
  we take this value to be as the typical value for long period variables
with thin envelopes (Knapp \& Morris 1985),
$\sim 5 \times 10^{-7} M_\odot \yr^{-1}$.
Interestingly, this mass loss rate is $\dot M_h \sim 0.1 L/vc$, for an AGB
stellar luminosity of $3 \times 10^3 L_\odot$ and wind speed of
$v=10 \km \s^{-1}$; $c$ is the speed of light.
 We note that oxygen-rich stars have a median mass loss rate of
$\sim 0.1 L/vc$ (Knapp 1986).
We take, therefore, the halo mass loss rate to be
$\dot M_h=5 \times 10^{-7} M_\odot \yr^{-1}$.
\item We take the mass loss rates in the two first phases to have a ratio
\begin{equation}
\eta \equiv \frac{\dot M_s}{\dot M_h}.
\label{eta1}
\end{equation}
Following the discussion in previous sections, we assume that for
spherical PNs $\eta=1$ (no FIW in spherical PNs), while $\eta>1$ for
elliptical and bipolar PNs.
\item We assume that the mass in the shell that was formed from
the FIW, $M_s$, is the same for all PNs.
As discussed later, taking instead the duration of the FIW, $t_s$,
to be the same, will make spherical PNs even more difficult to be detected
relative to non-spherical PNs.
\item We neglect the different evolution time of central stars having
different masses (e.g., Bl\"ocker 1995) and take the total time the
central star ionizes the nebula to be $\Delta t_\ast =30,000-40,000 \yr$.
  This assumption, assumption (5), and assumption (8) above are
clearly not true for massive progenitors, many of which form bipolar
and some which form extreme elliptical PNs. As can be seen from Table (1),
this can introduce uncertainties of $\sim 15-25 \%$, still smaller
than some other uncertainties, considering that we argue for a hidden
population larger by an order of magnitude than the observed population.
In other words, the parameter most influential in the toy model is $\eta$.
\item We take most known PNs to be located in a slab, with the galactic
plane at its center, and we take a constant density of PNs per unit area,
$\sigma_{\rm PN}$.
Namely, we take a uniform two dimensional distribution of PNs, neglecting
the high density toward the galactic plane.
We also take most PNs to be in the distance range of
$1 \kpc \sim D_{\rm min} < D < D_{\rm max} \sim 5 kpc$.
\end{enumerate}
Despite being strong, these assumptions capture the basic factors involved
in a possible undetected population of spherical PNs.

The luminosity of the nebula $L$ is the sum of the halo and shell
luminosity.
The halo is bounded between $r_s=v t_s$ and $r_h=v t_h$,
and its density is $\rho_h=\dot M_h/(4 \pi r^2 v)$;
a similar expression holds for the shell bounded by $r_r=v t_r$
and $r_s=v t_s$.
The total mass in the halo is $M_h=\dot M_h (t_h-t_s)$, and that
in the shell $M_s=\dot M_s (t_s-t_r)$.
We find
\begin{eqnarray}
L= \int  k \rho^2 dV =
\frac{k}{4 \pi v^2} \left[
\dot M_s^2\left(\frac{1}{r_r}-\frac{1}{r_s} \right)
+ \dot M_h^2\left(\frac{1}{r_s}-\frac{1}{r_h} \right) \right]
\nonumber \\
= \frac{k}{4 \pi v^3}
\left[ \frac{\dot M_s M_s}{t_s t_r}+\frac{\dot M_h ^2 (t_h-t_s)}{t_h t_s} \right],
\label{L1}
\end{eqnarray}
where  $dV=4 \pi r^2 dr$ is a volume element, and $k$ is a constant
such that $l=k \rho^2$ is the emissivity (power per unit volume).
At a temperature of $10^4 \K$ we find for $H \beta$ that
$k=2.6 \times 10^{22} \erg \cm^3 \s^{-1} \g^{-2}$.
Using our assumption $t_h \gg t_s$, and the relation
$t_s=t_r+M_s/ \dot M_s$, equation (\ref{L1}) becomes
\begin{equation}
L=\frac{k}{4 \pi v^3}
\left( \frac{\dot M_s M_s}{t_r}+\dot M_h ^2 \right)
\left(\frac{M_s}{\dot M_s}+t_r \right)^{-1} .
\label{L2}
\end{equation}
Using the definition of $\eta$ (eq. \ref{eta1}), defining
\begin{equation}
\tau \equiv \frac {M_s}{\dot M_h}= 4 \times 10^5
\left( \frac{M_s} {0.2 M_\odot} \right)
\left( \frac{\dot M_h} {5 \times 10^{-7} M_\odot \yr^{-1}} \right)^{-1}
\yr,
\label{tau1}
\end{equation}
and
\begin{equation}
L_s \equiv \frac{k}{4 \pi v^3} \frac {\dot M_h^2}{\tau},
\label{Ls}
\end{equation}
we can write equation (\ref{L2}) as
\begin{equation}
L=L_s \left(1+\eta\frac{\tau}{t_r} \right)
\left(1+\frac{1}{\eta} \frac{\tau}{t_r} \right)^{-1}
\frac{\tau}{t_r}.
\label{L3}
\end{equation}

By our assumption (2), for detecting a planetary nebulae
$F=L/(4 \pi D^2) >F_0$, where $D$ is the distance to the PN.
We write this condition as
\begin{equation}
F=F_s  D_{\rm kpc} ^{-2}
\left(1+\eta\frac{\tau}{t_r} \right)
\left(1+\frac{1}{\eta} \frac{\tau}{t_r} \right)^{-1}
\frac{\tau}{t_r}>F_0,
\label{F1}
\end{equation}
where $D_{\rm kpc}$ is the distance to the PN in units of kpc, and
\begin{eqnarray}
F_s (H \beta) \equiv \frac{L_s}{4 \pi (1 \kpc )^2} \simeq
10^{-15}   
\left( \frac{\dot M_h}{5 \times 10^{-7} M_\odot \yr^{-1}} \right)^3
\left( \frac{\dot M_s}{0.2 M_\odot} \right)^{-1}
\nonumber \\
\times \left( \frac{v}{10 \km \s^{-1}} \right)^{-3}
\erg \s^{-1} \cm^{-2}
\label{Fs}
\end{eqnarray}
Solving for $t_r$, we find the time during which the
PN can be detected.
Equation (\ref{F1}) can be cast into a quadratic equation
\begin{equation}
\left( \frac{t_r}{\tau} \right)^2
+\left( \frac{t_r}{\tau} \right)
\left(\frac{1}{\eta}-\frac{F_s}{F_0} D_{\rm kpc}^{-2} \right)
-\frac{F_s}{F_0} \eta D_{\rm kpc}^{-2} < 0.
\label{tr1}
\end{equation}

Neglecting the post-AGB evolution time (assumption 5 above), the
time span during which a PN at distance D with enhanced mass loss
rate factor $\eta$ can be detected, is given by the solution
of the last equation.
We scale the time with $\tau$, and marked it by
$\Delta t(\eta, D) \equiv t_r/\tau$.
However, there is another condition on $\Delta t$.
In the treatment above we have neglected the fading of the ionizing
central star.
The condition we should add is that the PN life time will not exceed
the time it takes the central star to fade, which itself
depends on the central star mass.
Ignoring this in our toy model, assumption (11) above, we add therefore
the condition
\begin{equation}
\Delta t(\eta, D) \le \Delta t_c = \frac{\Delta t_\ast}{\tau} \simeq 0.1,
\label{deltat3}
\end{equation}
where the values used come from equation (\ref{tau1}) and assumption (11).
The last condition implies that for a distance $D<D_c(\eta)$ we should
set $\Delta t(\eta,D) = \Delta t_c$ from equation (\ref{deltat3}),
rather than taking the solution of equation (\ref{tr1}).
Solving equation (\ref{tr1}) and using condition (\ref{deltat3}), gives
the time span during which the PN can be detected
\begin{equation}
\Delta t(\eta, D) =
{\rm min} \left\{ \frac{1}{2} \left(\frac{F_s}{F_0} D_{\rm kpc}^{-2}-\frac{1}{\eta}\right)
+\frac{1}{2} \left[\left(
\frac{F_s}{F_0} D_{\rm kpc}^{-2}-\frac{1}{\eta}\right)^2
+4 \frac{F_s}{F_0} \eta D_{\rm kpc}^{-2}
\right]^{1/2} \quad , \quad \Delta t_c \right\}.
\label{deltat1}
\end{equation}
The case $\eta=1$ has a simple solution
\begin{equation}
\Delta t(\eta=1, D) = {\rm min} \left\{ \frac{F_s}{F_0} D_{\rm kpc}^{-2}
\quad , \quad \Delta t_c \right\}.
\label{deltat2}
\end{equation}

The critical distance $D_c$ below which $\Delta t_c$ should be used
is found by substituting $t_r/\tau=\Delta t_c$ in equation (\ref{tr1}).
This gives
\begin{equation}
D_c= \left[\frac{F_s}{F_0}\frac{\Delta t_c + \eta}{\Delta t_c}
\left(\Delta t_c +\frac{1}{\eta} \right)^{-1} \right]^{1/2}
\simeq
\left[(1+10 \eta)
\left(1+ \frac{10}{\eta} \right)^{-1} \right]^{1/2},
\label{dc1}
\end{equation}
where in the second equality we substituted the typical numerical
values used here, $F_s/F_0=0.1$ (see next subsection), and
$\Delta t_c =0.1$.

\subsection{Numerical Values}
To estimate the value of $\eta$ we start with the
density profile of the PN NGC 6826 as given
by Plait \& Soker (1990).
This PN has a large spherical Halo, with
an elliptical inner structure, similar
to several other PNs (Corradi et al. 2003).
For a constant mass loss rate the density decreases
with radius as $R^{-2}$.
In the boundary between the inner elliptical region
and the outer spherical halo the density gradient is
much steeper.
In going from $\sim 12$~arcsec to $\sim 18$~arcsec the density
decreases by a factor of $\sim 5$.
Comparing to $(18/12)^2=2.25$, we find that the increase
in mass loss rate was $\eta \sim 2.5$.
In a study of the proto-PN HD 56126 Meixner et al.\ (2004) find that the mass
loss rate during the superwind (FIW), which lasted 840 years, was $\eta =6$
times higher than that in the previous mass loss episode,
$\dot M = 3 \times 10^{-5} M_\odot \yr^{-1}$ versus
$\dot M = 5.1 \times 10^{-6} M_\odot \yr^{-1}$.
We therefore take $\eta=4$ to be our standard value.

The value of the detection flux limit is very hard to give, because
PNs are discovered by many different telescopes and surveys.
Our approach is to examine the histogram of $H\beta$ flux of
PNs from the catalogue of Acker et al.\ (1992; their fig. 9a).
The lower detection flux is $\sim 10^{-15} \erg \s^{-1} \cm^{-2}$,
with the majority of PNs with $F>10^{-14} \erg \s^{-1} \cm^{-2}$,
which we take to be the detection limit $F_0$.
Therefore, our toy-model value is
$F_s/F_0=0.1$, where $F_s$ is by equation (\ref{Fs}).

\subsection{Detection Probability}
The chance to detect a PN of a morphological type which we characterize
by the enhanced mass loss rate factor $\eta$, at a distance $D$, is the
fraction of time for which it obeys equation (\ref{F1}), which solution
is given by equation (\ref{deltat1}),
\begin{equation}
P_{\eta D}(D) \propto \frac {\Delta t(\eta,D)}{t_L(\eta)}
N_\eta(D),
\label{p1}
\end{equation}
where $t_L(\eta)$ is the life time of the PN, assumed not
to depend on $\eta$ in our toy model, and
$N_\eta(D)=2 \pi \sigma_{\rm PN} D dD$ is the number of PNs between
$D$ and $D+dD$, and we used our assumption (12) of
constant PN density per unit area in the disk.
Normalizing the probability under our assumption of PNs slab
distribution, gives
\begin{equation}
P_\eta = \frac{1}{\pi (D_{\rm max}^2-D_{\rm min}^2) \sigma_{\rm PN}}
\int^{D_{\rm max}}_{D_{\rm min}} 2 \pi D \sigma_{\rm PN}
\frac {\Delta t(\eta,D)}{t_L(\eta)} dD
\label{p2}
\end{equation}
As we are interested only in relative probabilities among the
different morphological types, marked by $\eta$ in our toy model,
we can cancel common factors, such as the life time $t_L$.
The relevant probability is then
\begin{equation}
P_{R\eta} \equiv \frac{P_\eta}{P(\eta=1)}=
\frac{1}{P_1}
\int^{D_{\rm max}}_{D_{\rm min}}  D
\Delta t(\eta,D) dD,
\label{p3}
\end{equation}
where with the aid of equation (\ref{deltat2}), we find
\begin{equation}
P_1 \equiv \int^{D_{\rm max}}_{D_{\rm min}}  D
\Delta t(\eta=1,D) dD=
\frac{F_s}{F_0} \ln \frac{D_{\rm max}}{D_{\rm min}}
\simeq 0.1 \ln D_{\rm max},
\label{p4}
\end{equation}
where in the second equality we substitute our standard values of
$D_{\rm min} =1$, and $F_s/F_0=0.1$.
For $\eta \gg 1$ such that $D_c \geq D_{\rm max}$,
\begin{equation}
P_{R\eta} (D_c \ge D_{\rm max}) =
\frac{1}{P_1}
\int^{D_{\rm max}}_{D_{\rm min}}  D
\Delta t_c dD =
\frac{ \Delta t_c (D_{\rm max}^2 - D_{\rm min}^2 )}{2}
\left( \frac{F_s}{F_0} \ln \frac{D_{\rm max}}{D_{\rm min}}\right)^{-1}
\simeq 7.46   ,
\label{p5}
\end{equation}
where in the second equality we substitute our standard values of
$\Delta t_c=0.1$, $D_{\rm min} =1$, $D_{\rm max} =5$, and $F_s/F_0=0.1$.
For $D_{\rm max} = 4$ and $6$, we obtain (for $D_c \ge D_{\rm max}$)
$P_{R\eta}= 5.4$, and $9.8$,  respectively.
In Figure 1 we draw $P_{R \eta}$ as function of $\eta$
for the above parameters.
A good fit to the line for the above parameters is
$P_{R\eta}(\eta<6.4) = 0.00682 \eta^4 - 0.10864 \eta^3 + 0.36082 \eta^2
+ 1.6704 \eta - 0.95117$, while for $\eta>6.4$, as we found above,
$P_{R\eta}(\eta \ge 6.4) = 7.46$.

When examining the results presented in Figure 1, one should
recall the purpose of this section.
We here do not prove and do not show that there is a large population
of undetected spherical PNs.
This claim (see previous section) is based on the small detection
fraction of spherical halos around elliptical and bipolar PNs
(Corradi et al. 2003), and our basic postulate that most spherical PNs
are similar in structure to spherical halos around non-spherical PNs.
In the present section we limit ourselves to show that it is indeed
possible that most spherical PNs are below detection limits of
previous observations.
In other words, in this section we show that there is no contradiction
between the claim that many spherical PNs are not detected, and the
known properties of PNs and halos of PNs, under our basic postulate.
For that, in this section we built a toy model,
and took reasonable values for the physical parameters.
Figure 1 shows just that, and no more. Namely, for our favored
toy model value for the mass loss enhancement factor $\eta=4$,
the detection probability of spherical PNs is an order of magnitude
below that of non-spherical PNs.
This value is only suggestive, because of the nature of the toy model,
but it compatible with the spherical-PN detection  fraction
of $\sim 10 \%$ found in section 3.

The toy model actually overestimates the detection probability of spherical
PNs relative to that of non-spherical PNs.
Our assumption (10) that all PNs have the same shell mass $M_s$, was
implemented by scaling the time with the same $\tau$ (eq. \ref{tau1}).
Lower FIW mass loss rate (lower $\eta$), implies lower density, but
wider shells. As the luminosity is proportional to density square
times the volume, the luminosity of the shell is proportional to $\eta$.
Had we taken the duration of the FIW that formed the shell to be
the same for all PNs, which implies the same shell's volume for all PNs,
the shell's luminosity would have been proportional to $\eta^2$.
Namely, the detection probability would have increases faster for higher
mass loss rate.
The same argument goes for a calculation to find the
detection probability by surface brightness.
Since the shell has the highest surface brightness, the detection of
PNs is by the shell, with negligible contribution from the halo
(the relative contribution of a large halo to the luminosity is
more significant than to the surface brightness).
The surface brightness of the shell goes as $\sim \eta^2$,
and not as $\eta$.
For these considerations we have build our toy model to be in
the safe side, in the sense that using other reasonable assumption
(like considering detection by surface brightness)
will only strengthen our claim for hidden population of
spherical PNs.

\section{SUMMARY}
The goal of the present paper is to update the statistical distribution
of different morphological types of PNs and their progenitors,
in the frame of the binary-shaping model.
This update is motivated by, and uses, recent results on the
distribution in mass and orbital separation of extra-solar planets,
and the new work of Corradi et al. (2003) on halos around PNs.
We make two assumptions:
\begin{enumerate}
\item We take the common view that the axisymmetric (including
point symmetric) structure of PNs result from binary interaction;
the companion can be stellar or substellar.
\item We assume that spherical PNs (according to the definition
of Soker 1997 which is used here) are similar in structure to the
spherical halos around non-spherical PNs.
  This assumption is based on the results of Soker (2002).
\end{enumerate}
Only the spherical morphological type is studied here.
We conclude that most probably there is a large population
of spherical PNs, much more than non-spherical PNs, that are
too faint to be detected.
This leads to a new estimate regarding the distribution
of PN progenitors and different PN morphology types
(summarized in sections 1 and 2).
The new estimates compared with previous ones
are summarized in Table 1.

To reach the conclusion that there is a hidden spherical PN
population, we first examine Corradi's et al. (2003) observations
and analysis of halos around PNs.
Whereas all PNs are expected to have a faint halo,
the detection fraction is much lower;
taking the PNs listed by Corradi et al. (2003) to have detectable
halos, and compare with all PNs, we find that only $\sim 12 \%$
of all PNs have detectable halos.
Using our second assumption above, this suggest that the detection
fraction of non-spherical PNs is $\sim 8$ times higher than that of
spherical PNs.
Because many of the PNs listed by Corradi at al. (2003) were a target
of specific (deeper) observation, the average (among all large surveys)
detection fraction of non-spherical PNs can be $\sim 10-15$ times
higher than that of spherical PNs.
We therefore claim that while $\sim 3 \%$ of all well resolved PNs
are spherical (Soker 1997), the true fraction of spherical
PNs among all PNs is likely to be $\sim 30 \%$.
The recent detection of a spherical PN at a distance of only
$\sim 0.5 \kpc$  (Frew \& Parker 2006) show that spherical and faint
PN can indeed escape detection.
This conclusion regarding a {\it hidden} population is not to be
confused with the claim of Manchado et al.\ (2000) for $\sim 25 \%$
{\it observed} spherical PNs; they simply use a different definition
(which we don't accept; see section 1) for spherical PNs.

We then (section 4) built a toy model to show that such a large
(relative to that of non-spherical PNs) hidden population of spherical
PNs might be compatible with present sensitivity of observations.
We stress here that in section 4 we do not determine or estimate the
detection fraction of spherical PNs.
The uncertainties in many physical parameters (in particular in the
progenitor's mass loss history) are too large to allow such an estimate.
However, the toy model shows that the claim for a large hidden spherical
PN population (section 3) is compatible with what we know about PNs.

  One interesting conclusion from this paper concerns the fraction
of PNs that were shaped by planetary systems.
As seen from Table 1, the required fraction of PNs that were shaped
by planets was reduced substantially from that estimated by Soker (1997).
The stellar population synthesis studies yield a relatively secure fraction
for the PN shaped by stellar companions.
The three stellar evolutionary routes add up to $\la 60\%$.
The rest of PN progenitors either yield a spherical PNs or
were interacting with substellar objects to form elliptical PNs.
In the past, most of these progenitors were assumed to have substellar
companions that influenced the mass loss geometry.
The present study reduces this number.
This is in accord with the finding that the fraction of Sun-like stars that
have close and massive planets or brown dwarfs around them is $\la 25 \%$
(see section 2).
We actually argue that the number of AGB stars that formed spherical PNs is larger
than those having substellar companions that influenced their mass loss process.
Our best estimate in the present study is that the hidden population of spherical
PNs, although large, is too small to account for all PNs not shaped by
stellar companions; hence shaping by planetary system seemed to be required
in some cases.
However, we cannot completely rule out the possibility that the
hidden population of spherical PNs is larger, and that practically no PN
was shaped by planets (some may have been shaped by brown dwarfs).

The hidden population of spherical PNs, if exist, will not change
much the comparison between the total number of PNs, and the number
expected from stellar counts. This is because the proposed hidden population
adds only $\sim 30 \%$ to the total number of PNs, a number much smaller
than the uncertainties in both population synthesis and in estimating
the total number of PNs in the galaxy.
  This is the place to stress again that our finding regarding the
hidden population of spherical should be considered in the context of the
general estimate of PN progenitor population presented in Table 1.
For example, the hidden population is compatible with the finding that
the fraction of stars having brown dwarfs or massive enough planets
to shape the descendant PNs is $\sim 25 \%$, but only a fraction of
them (estimated here to be $\sim 15 \%$ will form PNs (see section 2).

The main prediction of this paper is that {\it very} deep observations
will detect more spherical PNs. This is not easy. After all, instead of
the 14 spherical PNs classified by Soker (1997), we expect that for
the same sample of PNs there will be $\sim 150$ spherical PNs,
which is a small number for deep observations of particular patches
of the sky
(they will not be highly concentrated in the galactic plane).
In addition, they are expected to be very faint, hence
to be detected they require long observations.
A survey such as the AAO/UKST H$\alpha$ one (Parker et al.\ 2003), might
discover higher percentage of spherical PNs.
Indeed, a circular shell around an evolved star that
was detected by this survey (Cohen et al.\ 2005).
However, this survey is based on surface brightness.
As discussed at the end of the previous section, in
surface brightness surveys spherical PNs are even less likely
to be discovered compared with non-spherical PNs.
Therefore, it is not clear whether the proposed hidden spherical PNs
population will be revealed by that search.


\acknowledgments
We thank Letizia Stanghellini, Romano Corradi, and Joel Kastner
for useful comments.
During early stages of this research we benefited from discussions
with Oren Schonberger.
This research was supported in part by a grant from the Israel
Science Foundation.


\begin{deluxetable}{ccccc}
\tablewidth{0pt}
\tablecaption{BINARY INTERACTION AND MORPHOLOGY OF DESCENDANT PNs \label{tab:binary}}
\tablehead{
\colhead{Binary interaction} & \colhead{Major PN type} & &
\colhead{Percentage} &  \\
\colhead{} & & \colhead{1997} & \colhead{2001} & \colhead{This paper} }
\startdata
Close stellar companion  &  &  &   & \\
outside envelope         & Bipolar &$\sim 11$& $\sim 15$& $\sim 15$ \\
\hline
Stellar companion in     & Extreme &  &    & \\
a common envelope        & Elliptical & $\sim 23$& $\sim 25$& $\sim 25$\\
\hline
Substellar companion   &    & & & \\
in a common envelope   &  Elliptical  &$\sim 56$&$\sim 35$& $\sim 15$\\
\hline
Wide stellar companion   & Elliptical+jets &$-$ & $\sim 15$& $\sim 15$\\
\hline
No interaction & Spherical   & $\sim 10$ & $\sim 10$ & $\sim 30$ \\
\enddata

\tablecomments{Percentage of the different morphological types and their
progenitors in the binary shaping model.
The second column gives the major,
but not the sole, PN descendant from the mode of binary interaction.
Percentage are of all PNs, and not of the observed PNs only.
As discussed in the paper, most spherical PN are too faint to
be detected.
The percentage are from Soker (1997), Soker (2001a,b),
and present paper.}

\end{deluxetable}

\begin{figure}
\includegraphics[width =135mm]{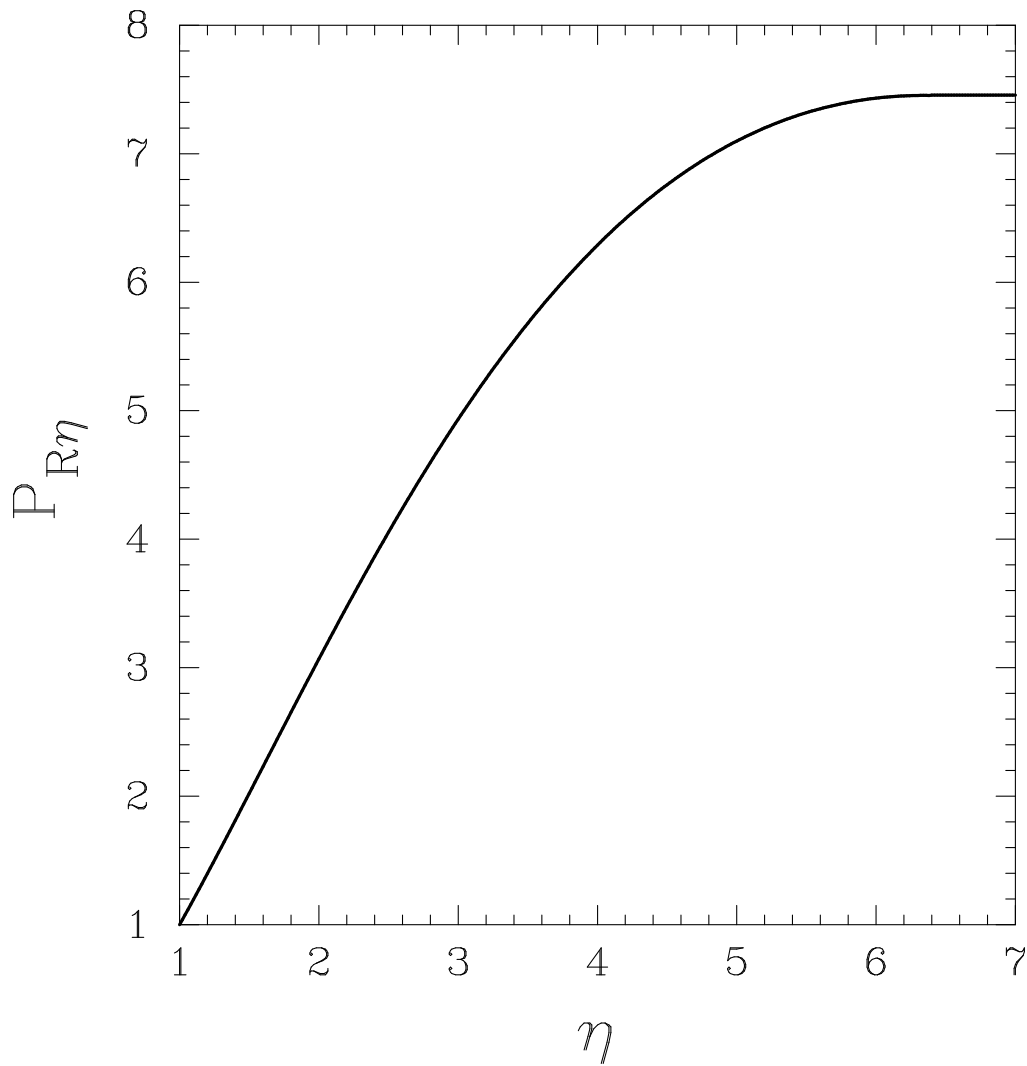} \vskip 0.2 cm
\caption{The probability of detecting a PN in the distance
range $1-5 \kpc$, as a function of the enhanced mass loss rate
factor $\eta$ (eq. \ref{p3}).
In the toy model, $\eta$ is the ratio between mass loss rate during
the final intensive wind (superwind) on the AGB and the regular AGB wind
(eq. \ref{eta1}).
The probability $P_{R \eta}$ is given relative to the probability of
detecting a PN with no enhanced mass loss rate, which is assumed to
be spherical. }
\end{figure}
\end{document}